\documentclass[sigconf]{acmart}

\usepackage{subfigure}
\usepackage{booktabs} 

\usepackage{mathrsfs}
\usepackage{enumitem}
\usepackage{epstopdf}
\usepackage{multirow}
\usepackage{graphicx}
\usepackage{balance}

\setcopyright{rightsretained}

\makeatletter
\def\hlinew#1{%
	\noalign{\ifnum0=`}\fi\hrule \@height #1 \futurelet
	\reserved@a\@xhline}

\begin{document}
	
	\copyrightyear{2018} 
	\acmYear{2018} 
	\setcopyright{acmcopyright}
	\acmConference[KDD'18]{The 24th ACM SIGKDD International Conference on Knowledge Discovery \& Data Mining}{August 19--23, 2018}{London, United Kingdom}
	\acmBooktitle{KDD'18: The 24th ACM SIGKDD International Conference on Knowledge Discovery \& Data Mining, August 19--23, 2018, London, United Kingdom}
	\acmPrice{15.00}
	\acmDOI{10.1145/3219819.3220014}
	\acmISBN{978-1-4503-5552-0/18/08}
	
	\title[KDD 2018 Research Paper]{Learning from History and Present: Next-item Recommendation\\ via Discriminatively Exploiting User Behaviors}

	\author{Zhi Li$^{1,2}$, Hongke Zhao$^{1}$, Qi Liu$^{1,*}$, Zhenya Huang$^{1}$, Tao Mei$^{3}$, Enhong Chen$^{1}$}
	\thanks{$^{*}$The corresponding author.}
	\affiliation{%
		\institution{
			$^{1}$Anhui Province Key Laboratory of Big Data Analysis and Application, University of Science and Technology of China;
			$^{2}$School of Software Engineering, University of Science and Technology of China;
			$^{3}$JD AI Research}
		\{zhili03,zhhk,huangzhy\}@mail.ustc.edu.cn; \{qiliuql,cheneh\}@ustc.edu.cn; tmei@jd.com
	}

	
	
	
	
	

	\settopmatter{printacmref=false, printccs=true, printfolios=true}
	
	\renewcommand{\shortauthors}{KDD 2018 Research Paper}

	\begin{abstract}
		In the modern e-commerce, the behaviors of customers contain rich information, e.g., consumption habits, the dynamics of preferences. Recently, session-based recommendations are becoming
		popular to explore the temporal characteristics of customers' interactive behaviors. However, existing works mainly exploit the short-term behaviors without fully taking the customers' long-term stable preferences and evolutions into account. In this paper, we propose a novel Behavior-Intensive Neural Network (BINN) for next-item recommendation by incorporating both users' historical stable preferences and present consumption motivations. Specifically, BINN contains two main components, i.e., \emph{Neural Item Embedding}, and \emph{Discriminative Behaviors Learning}. Firstly, a novel item embedding method based on user interactions is developed for obtaining an unified representation for each item. Then, with the embedded items and the interactive behaviors over item sequences, BINN discriminatively learns the historical preferences and present motivations of the target users. Thus, BINN could better perform recommendations of the next items for the target users. Finally, for evaluating the performances of BINN, we conduct extensive experiments on two real-world datasets, i.e., Tianchi and JD. The experimental results clearly demonstrate the effectiveness of BINN compared with several state-of-the-art methods.
		
	\end{abstract}
	
	%
	%
	


	\maketitle
	
	\section{Introduction}
	
	\begin{figure}[tbp]
		\vspace{+0.1in}
		\centering
		\includegraphics[width=3.3in,height=1.8in]{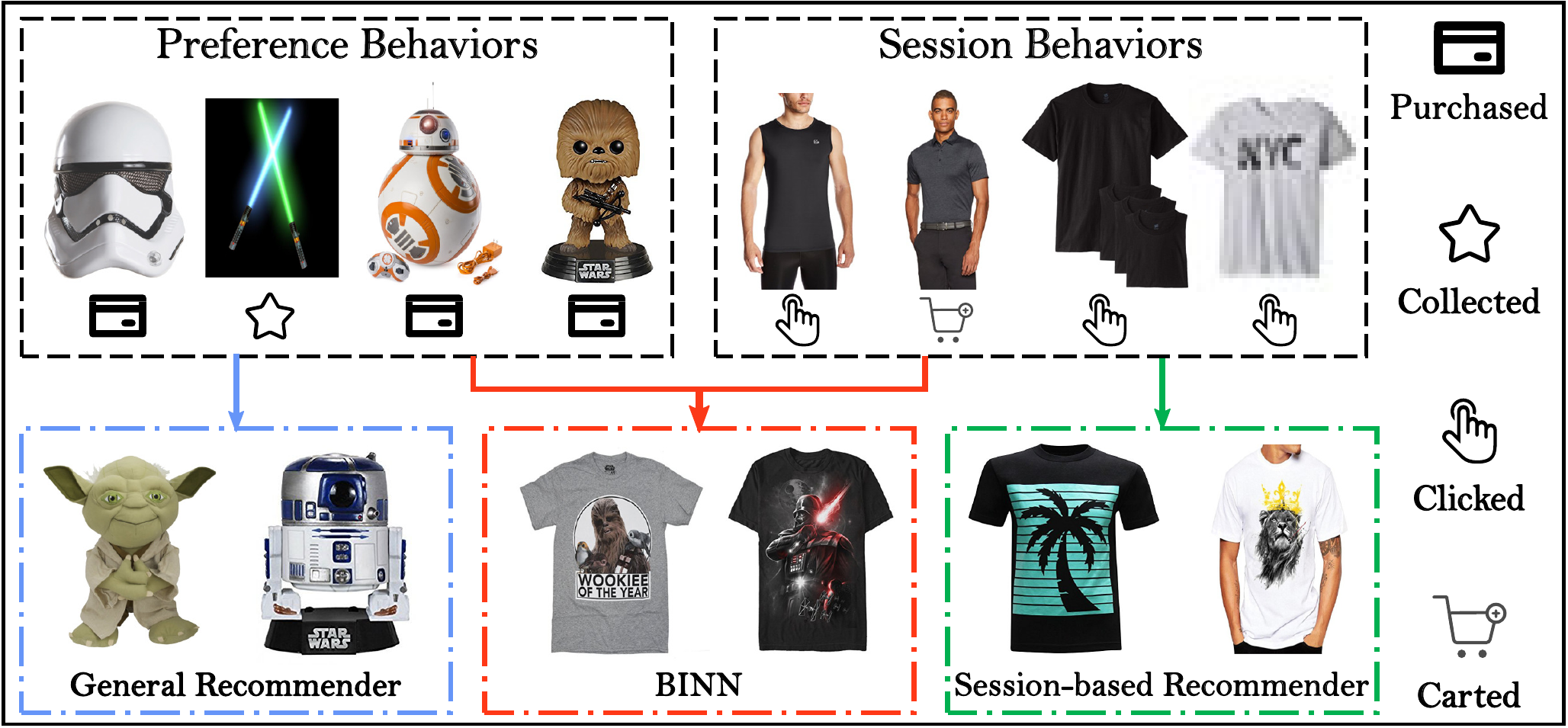}\vspace{-0.1in}
		\caption{Recommending by integrating preference behaviors and session behaviors.}\vspace{+0.1in}
		\label{case}
	\end{figure}
	
	Recommender system, as an essential component of modern e-commerce websites, tries to predict what the most suitable products or services are of users, based on the users' preferences~\cite{ricci2015recommender}. With the mechanism development of e-commerce, a massive amount of customer interactions (e.g., browse, click, collect, cart, purchase) have been logged, which imply luxuriant consumption patterns. These information-rich logs provide opportunities for understanding customers' historical stable preferences and also their present consumption motivations, which may further contribute to smarter recommendations.

	Along this line, there is a particular interest in understanding interactive behaviors of customers. Existing works can be concluded into two main paradigms. The first paradigm is the general recommenders. These works focus on mining the static relevancy between users and items from interactions, which are represented by the traditional collaborative filtering models~\cite{koren2010collaborative, liu2015mining, zhang2016collaborative}. For example, Zhang \emph{et al.} made recommendations through a factorization model with different item semantic representations from the knowledge base~\cite{zhang2016collaborative}. However, most of these works have taken user-item specific relationships into consideration from the static views but neglect the dynamics and evolutions of users' preferences implied in sequential interactions. The other paradigm is recommending next items based on sequential pattern mining~\cite{yap2012effective,shang2014personalized} or transition modeling~\cite{rendle2010factorizing,zhao2017sequential}. Along this line, researchers recently show more interest in an e-commerce scenario where user profiles are invisible so that recommender systems are developed based on the user interactions in short sessions~\cite{hidasi2015session, Quadrana:2017:PSR:3109859.3109896, Li:2017:NAS:3132847.3132926}. These session-based models have provided the comprehension about users' decision-making process in a short term, but the dynamics of preferences~\cite{wu2016unfolding} and how to perfectly integrate both the historical stable preferences with present consumption motivations are still largely unexplored.

	Actually, as a user's interactive behaviors naturally form a behavioral sequence over time, the user's historical preferences from the long-term view and present motivations or demands from the short-term view can be dynamically revealed.
	For instance, Figure~\ref{case} illustrates a typical online shopping scenario. 
	The user's historical interactions imply that this user might be a ``Star Wars'' fan since that the user has bought or collected various spin-off products of the ``Star Wars''. Moreover, we infer that this user would like to buy dark T-shirts because a black shirt is included in the personal cart and the user has browsed many short sleeve shirts. However, following the general collaborative filtering approaches, another spin-off product may be recommended since all the preference behaviors of the user's entire history are exploited in the static manners as shown in the blue chart of Figure~\ref{case}. By contraries, if we only consider the current session behaviors of this user in accordance with what the session-based models do, another similar or popular shirt would be recommended as shown in the green chart of Figure~\ref{case}. Actually, by exploiting both the historical stable preferences and present consumption motivations of this user, more attention should be paid to short sleeve shirts and the ``Star Wars'' graphic T-shirts perfectly match user's tastes. Therefore, we can conclude that a smarter recommender system should not only consider users' historical stable preferences but also take into account the present consumption motivations by discriminatively exploiting different terms or types of user behaviors.

	Based on the intuition and observations, we propose a novel solution framework called Behavior-Intensive Neural Network (BINN) to address the next-item recommendation problem. Our BINN framework contains two main components: \emph{Neural Item Embedding} and \emph{Discriminative Behaviors Learning}. Specially, we propose a novel neural item embedding method to obtain a unified item representation space for learning latent vectors which could capture the sequential similarities between items. Different from the traditional item embedding methods which are based on inherent features such as item images or text descriptions, our neural item embedding method generates item representations by means of exploiting users' collaborative sequential interactions over items directly. Then, with the item embedded, we design two different behavior alignments, i.e., \emph{Session Behaviors Learning} and \emph{Preference Behaviors Learning} to respectively model users' present consumption motivations and historical stable preferences by discriminatively exploring interactive behaviors of users. Specific to the alignments, we respectively develop two deep neural network architectures to jointly learn the session behaviors and preference behaviors. Finally, by matching the potentially preferred items in the latent space, BINN generates recommendations for the target users. For evaluating the performances of BINN, we conduct extensive experiments on two real-world datasets. The experimental results clearly demonstrate the effectiveness of BINN compared with several state-of-the-art methods. In summary, the main contributions of this study can be summarized as follows.
	
	\begin{itemize}[leftmargin=*,itemsep=2.5pt]
		\item We propose to make item-recommendation by integrating both the historical preferences and present motivations of users, which are all learned from the users' interactive behaviors.
		
		\item We propose a novel Behavior-Intensive Neural Network (BINN) which includes embedding items by users' interactions and discriminative behavior alignments accompanied by two applicable neural network architectures.
		
		\item We conduct extensive experiments on two real-world datasets. The results show that BINN model outperforms other state-of-the-art methods from various aspects.
	\end{itemize}

	\section{Related Works}
	Recommender systems play an essential role in many Internet-based services~\cite{ricci2015recommender}, such as e-commerces, and have arouse the great attention from both industry and academia. The relevant works of this study can be concluded into two main paradigms: the \emph{General Recommenders} and the \emph{Sequential Recommenders}.

	\subsection{General Recommenders}
	Most of the general recommenders focus on mining the static relevancy between users and items from their interactions, which are represented by the traditional collaborative filtering models~\cite{koren2010collaborative, liu2015mining, mnih2008probabilistic}. More specifically, the most common approaches of collaborative filtering are the \emph{neighborhood methods}~\cite{sarwar2001item,linden2003amazon, bell2007scalable, koren2008factorization, zhao2014investment} and the \emph{factorization models}~\cite{cui2011should, mnih2008probabilistic, he2016fast, He:2017:NCF:3038912.3052569, zhao2016group}.
	
	\textbf{Neighborhood Methods} are based on the similarities of entities (always users or items) and recommend the nearest neighbor items through the precomputed similarities~\cite{sarwar2001item, bell2007scalable, koren2008factorization, zhao2014investment}. For example, Bell \emph{et al.} proposed a new method to simultaneously derive interpolation weights for all nearest neighbors, leading to a substantial improvement of prediction accuracy~\cite{bell2007scalable}. Zhao \emph{et al.} considered the expertise of investors to improve the neighborhood methods for the personalized investment recommendations in P2P lending~\cite{zhao2014investment}. Wang \emph{et al.} presented a basic probabilistic framework which formalized the learning similarity as a regression problem~\cite{wang2016probabilistic}. Moreover, they introduced a novel multi-layer similarity descriptor which modeled the joint influences of different features to improve the neighborhood methods~\cite{wang2016probabilistic}.
	
	\textbf{Factorization Models} treat the recommendation as a user-item matrix reconstruction problem and model the user-item interactions by dot product of latent vectors~\cite{mnih2008probabilistic, liu2011personalized, zhao2016group}. Many previous works focused on better representing users or items to improve the qualities of recommender systems. For example, Liu \emph{et al.} considered users' indecisiveness when the customers chose among competing product options and then proposed IMF method to mine the indecisiveness in customer behaviors to improve the performance of recommendation~\cite{liu2015mining}. Zhao \emph{et al.} introduced the Nash Equilibrium into the matrix factorization method to solve the group recommendation problem~\cite{zhao2016group}. In recent years, deep learning has also been applied successfully to the classical collaborative filtering user-item matrix reconstruction problems from different perspectives. For example, many researches incorporated deep learning models for extracting and fusing side features, such as image features~\cite{AAAI1611914}, text information~\cite{Elkahky:2015:MDL:2736277.2741667, Shang2012User} and multimode data~\cite{zhang2016collaborative}. Moreover, some works developed the deep neural networks to learn the relationships between the users and items~\cite{He:2017:NCF:3038912.3052569, ijcai2017-239}.
	
	However, most of these works provide recommendations by mining the static relevancy between users and items. The dynamics and evolutions of users' preferences, and also their present consumption motivations are usually not given special attentions.

	\subsection{Sequential Recommenders}
	In recent years, researchers start to focus on various sequential recommendation scenarios, such as next-basket~\cite{rendle2010factorizing, Wang:2015:LHR:2766462.2767694}, session-based~\cite{hidasi2015session, Quadrana:2017:PSR:3109859.3109896, Li:2017:NAS:3132847.3132926} and next-item recommendations~\cite{yap2012effective, Donkers:2017:SUR:3109859.3109877}. Among them, session-based ones are increasingly attractive.

	Early works on sequential recommenders are almost based on the sequential pattern mining~\cite{yap2012effective, shang2014personalized} and the transition modeling~\cite{rendle2010factorizing,zhao2017sequential}.
	Due to the tremendous success of deep neural networks in the past few years, approaches to sequence data modeling have made significant strides and benefit a broad range of applications, such as NLP~\cite{ghosh2016contextual}, social media~\cite{wu2017sequential} and recommendations~\cite{hidasi2015session, Quadrana:2017:PSR:3109859.3109896, Li:2017:NAS:3132847.3132926}. For example, Hidasi \emph{et al.} firstly applied recurrent neural networks (RNNs) by modeling the whole session for session-based recommendations, which outperformed item-based methods significantly~\cite{hidasi2015session}. Quadrana \emph{et al.} focused on some session-based recommendation scenarios where user profiles were readily available and developed a hierarchical recurrent neural networks with cross-session information transfer~\cite{Quadrana:2017:PSR:3109859.3109896}. Wang \emph{et al.} used a hierarchical representation model to capture both sequential behaviors and users' general tastes by involving transaction and user representations in next-basket prediction~\cite{Wang:2015:LHR:2766462.2767694}. Donkers \emph{et al.} extended RNNs by representing the individual users for the purpose of generating personalized next item recommendations~\cite{Donkers:2017:SUR:3109859.3109877}.
	
	Although these models have taken the users' sequential information into consideration, there are still largely unexploited in the coherence of customers' sequential behaviors and the dynamics of historical preferences. Comparatively, in this paper, we propose to make item-recommendation by integrating both the historical preferences and the present motivations of users, which are all learned from the users' rich interactive behaviors over items.

	\begin{figure*}
		\centering
		\includegraphics[width=7.0in,height=2.8in]{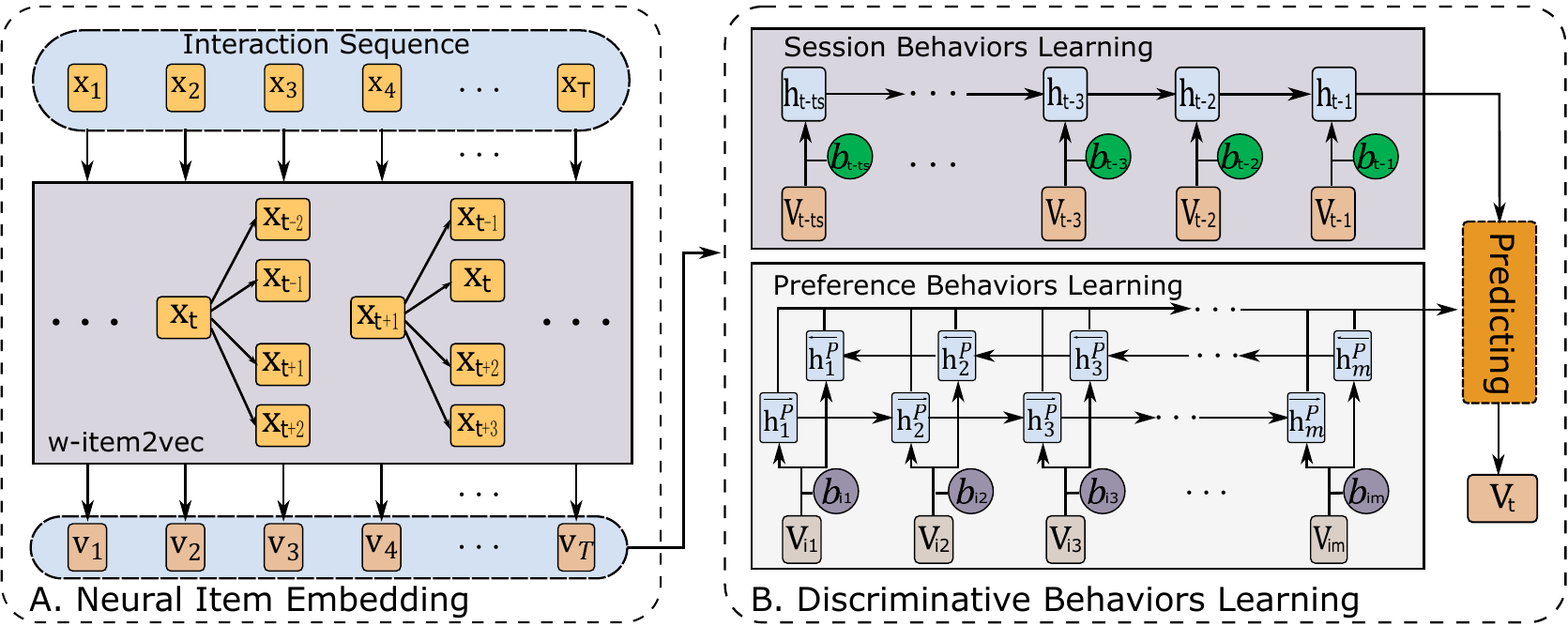}
		\caption{The overview of Behavior-Intensive Neural Network (BINN). A. Neural Item Embedding converts sequential items into a unified embedding space by w-item2vec. B. Discriminative Behaviors Learning constructs two alignments of user behaviors and discriminatively learns behavior information based on two LSTM-based architectures.}
		\label{framework}\vspace{-0.2in}
	\end{figure*}
	
	\section{BINN: Behavior-Intensive Neural Network}
	In this section, we introduce our proposed model for addressing the personalized next-item recommendations. We first formally define the next-item recommendation task and overview the proposed \emph{Behavior-Intensive Neural Network} (BINN). Then we describe the two main components of BINN in detail, i.e., \emph{Neural Item Embedding} and \emph{Discriminative Behaviors Learning}.
	
	\subsection{Preliminaries}
	Next-item recommendation is the task of predicting what a user would like to access next based on her historical interactions. Here, we give a formulation for the next-item recommendation.
	
	As user interactions naturally form a sequence over time, a log history $H$ of the information system is a set of sequential interactions, i.e., $H=\{S_1, S_2, ..., S_n\}$, where $|H|=n$ denotes the number of users. Each user $u$ has a corresponding interaction sequence $S_u \in{H}$ which can be represented as $S_{u}=\{(x_1, b_1), (x_2, b_2), ..., (x_T, b_T)\}$, where $x_j$ denotes the $j$-th item that user $u$ operates and $b_j$ denotes the behavior type (e.g., click, cart, purchase). Then, the personalized next-item recommendation task can be defined as follow:
	
	\vspace{+0.05in}
	\textbf{Definition 1} {(\textbf{Personalized Next-item Recommendation}). Given a target user $u$ with her sequential of interactive behaviors over items $S_{u}=\{(x_1, b_1), (x_2, b_2), ...,(x_T, b_T)\}$ and also all users' sequential interactions $H$, the personalized next-item recommendation task is to predict item $x_{T+1}$ that the target user $u$ is most likely to access in her next visit.}
	\vspace{+0.05in}

	In this paper, we address this task with a novel personalized next-item recommendation framework, i.e., Behavior-Intensive Neural Network (BINN). As shown in Figure~\ref{framework}, BINN contains two main components: Neural Item Embedding and Discriminative Behaviors Learning. Specifically, with a neural embedding model, we obtain a unified item representation space for learning the latent vectors that capture sequential similarities between items. Then we design two interactive behaviors assignments, named \emph{Session Behaviors Learning} (SBL) and \emph{Preference Behaviors learning} (PBL), to exploit the users' present consumption motivations and historical stable preferences over time. Finally, we jointly learn these two interactive alignments on the latent space of item representations and recommend top-k potentially preferred items to each target user.
	
	\subsection{Neural Item Embedding} \label{section:embedding}
	In the first stage of BINN, Neural Item Embedding aims to generate a unified representation for each item by learning the item similarities from a large number of sequential behaviors over items. Previous works of sequential recommenders always use 1-of-N encoding or add an additional embedding layer in the deep learning architecture to represent items~\cite{hidasi2015session, Quadrana:2017:PSR:3109859.3109896}. However, for a superb collection of items in the big e-commerce platforms, on one hand, the 1-of-N encoding networks may cost unaffordable time and always cannot to be optimized well because of the high sparsity~\cite{bengio2013representation}. 
	On the other hand, adding an additional embedding layer may make networks lose performances to some extend~\cite{hidasi2015session}. What's more, both two methods cannot reveal item sequential similarities which is implied in the user interactions. In this case, it is necessary to find an effective representation method to directly learn high-quality item vectors from the users' interaction sequences, with the result that items implied similar attractions tend to be close to each other.
	
	In recent years, the progress in neural embedding has achieved tremendous advances in many domains, such as NLP~\cite{mikolov2013distributed, bengio2013representation}, social networks~\cite{perozzi2014deepwalk, grover2016node2vec,cui2017survey} and recommendations~\cite{barkan2016item2vec}. Among these works, item2vec~\cite{barkan2016item2vec} is one of the significant extensions of Skip-gram with Negative Sampling~\cite{mikolov2013distributed} to produce item embedding for item-based collaborative filtering recommendations.

	In this paper, we propose an improved item2vec to capture item similarities and generate item representations by the means of exploiting users' collaborative interactions over items directly.
	Different from the words in sentences, some items in user interactions have often been frequently accessed. The reason for this phenomenon is that users are usually indecisive in their decision-making process~\cite{liu2015mining}, causing a lot of repetitive actions on the same item. In addition, these frequently-operated items also indicate users' main motivations, and other items interspersed these repeats may be very similar or competing to these items. On the other hand, items with low frequency may be aimlessly clicked~\cite{Li:2017:NAS:3132847.3132926}, which will bring noise to the item embedding. Along this line, we take one step further to capture the characteristic of interaction sequences and propose a novel item embedding method, called \emph{w-item2vec}, which considers the frequency of items as a weighted factor. 
	
	Inspired by item2vec~\cite{barkan2016item2vec}, w-item2vec also uses a Skip-gram model with Negative Sampling method~\cite{mikolov2013distributed}. Specifically, given an item sequence $S_u^{'}=\{x_1, x_2, ..., x_N\}$ of user $u$ from the interactive sequence $S_u$, the Skip-gram of w-item2vec aims at maximizing the following objective:
	
	\begin{equation}
	\arg\max \enskip opt = \frac{1}{K}\sum_{i=1}^{K}\sum_{j\neq{i}}^{K}\emph{log}~p(x_j|x_i),
	\label{skipgram}
	\end{equation}
	
	\noindent where $K$ is the length of sequence $S_u$, and $p(x_j|x_i)$ is defined as the softmax function:
	
	\begin{equation}
	p(x_j|x_i)=\frac  {exp(w_i^T*v_j)}{\sum_{k}exp(w_i^T*v_k)},
	\label{softmax}
	\end{equation}
	
	\noindent where $w_i$ and $v_i$ are the latent vectors that correspond to the target and context representations for item $x_i$. For alleviating computational complexity of the gradient $\bigtriangledown(x_j|x_i)$, Eq.~\eqref{softmax} is always replaced by negative sampling:
	
	\begin{equation}
	p(x_j|x_i)=\sigma(w_i^T*v_j)\prod_{k=1}^{E}\sigma(-w_i^T*v_k),
	\label{NS}
	\end{equation}
	
	\noindent where $\sigma(x)=1/(1+exp(-x))$, $E$ is the number of negative samples to be drawn per a positive sample.
	
	Then, we improve the negative sampling model of Eq.~\eqref{NS}{\footnote{There is a mistake in the version of ACM Digital Library, we correct it here.}} by considering the item frequency within interaction sequences as the weight of negative sampling process: 
	
	\begin{equation}
	p(x_j|x_i)=(\sigma^{\Theta(x_i)}(w_i^T*v_j) \prod_{k}\sigma(-w_i^T*v_k))^{\Theta(x_i)},
	\label{NSw}
	\end{equation}
	
	\noindent where $\Theta(x_i)$ is the frequency of item $x_i$ in the sequence. Consequently, the Skip-gram objective in Eq.~\eqref{skipgram} can be redefined as:
	
	\begin{equation}
	\begin{split}
	\arg \max \enskip opt &= \frac{1}{K}\sum_{i=1}^{K}\sum_{j\neq{i}}^{K}\emph{log}~p(x_j|x_i)\\
	&= \frac{1}{K}\sum_{i=1}^{K}\sum_{j\neq{i}}^{K}{\Theta(x_i)}(\Theta(x_i)*\emph{log}~\sigma(w_i^T*v_j)\\
	&\quad+{\sum_{k=1}^{E}}\emph{log}~\sigma(-w_i^T*v_k)).
	\end{split}
	\label{witem2vec}
	\end{equation}
	
	Finally, we train the w-item2vec by gradient descent, and obtain high-quality distributed vector representations for all the items.
	
	With w-item2vec, we can capture item similarities with the help of user interactions and generate an unified item representation space, in which the representation vectors can reveal similarities and sequential relationships of items. And for each user $u$, we can generate an interaction sequence with embedding items as $P_u=\{v_1, v_2, ..., v_T\}$, where $v_j$ denotes the $d$-dimensions latent vector of item $x_j$.

	\subsection{Discriminative Behaviors Learning}
	After obtaining item embeddings, Discriminative Behaviors Learning (DBL) could explore sequential behaviors as prior knowledge to recommend items that the target user is most likely to access in her next visit.
	
	As illustrated in Figure~\ref{case}, a user's decision-making process is mainly influenced by two factors: her present motivations and historical preferences. More specifically, users' present consumption motivations are dynamic in a short term and the recent fluctuations are also important to reflect the short-term characteristics. Considering that all the recent behaviors (e.g., click, collect, cart, purchase) may imply users' present motivations in a short term, we use all types of recent behaviors to represent the present consumption motivations. On the other hand, as for exploiting users' historical preferences, not all types of behaviors could depict users' preferences. For example, we can imply that the user do not prefer an item if she just clicks on it without purchasing at last. Therefore, for modeling users' historical preferences, we only remain the behaviors which can clearly depict users' underlying preferences from interaction histories, i.e., purchase behaviors.
	
	In fact, the interaction process of a user is the series of implicit feedbacks over time. Thus, different from traditional recommender systems which explore the user-item interactions from a static manner, we tackle the next-item recommendation by sequential modeling. Specifically, we design two discriminative behavior alignments: Session Behaviors Learning (SBL) and Preference Behaviors Learning (PBL) to discriminatively learn users' present consumption motivations and historical stable preferences. Further, on this basis, we develop two separate deep neural architectures based on LSTM~\cite{hochreiter1997long} to jointly learn the motivations and preferences from these two alignments of behaviors.

	\subsubsection{Session Behaviors Learning}
	As illustrated in the green chart of Figure~\ref{case}, the session behaviors in a short term can reveal the users' present consumption motivations.
	The Session Behaviors Learning (SBL) is to model the short-term session behaviors of the target user $u$. More formally, suppose we already have the previous interaction sequence $S_{u}=\{(x_1, b_1), (x_2, b_2), ..., (x_{t-1}, b_{t-1})\}$ with embedding items $P_u=\{v_1, v_2, ..., v_{t-1}\}$. The behavior $b_i$ can be represented as an one-hot vector and the length of the vector is the number of interaction types, e.g., click. For determining whether a certain item ${x_i}$ would be a possible element of the session behaviors, the SBL discrimination function $D_{SBL}$ can be defined as follows:

	\begin{equation}
	D_{SBL}(x_i,x_t)=\Phi{((t-i)\leq{ts})},
	\end{equation}
	
	\noindent where function $\Phi(a)$ is to compute a discrimination signal that equals to 1 if $a$ is true and equals to 0 otherwise, $x_{t-1}$ is the previous item of the prediction and $ts$ is a controlling indicator to control the length of SBL. Specifically, as for a session-based scenario, $ts$ is the length of the session. And for the non-session scenarios, $ts$ is artificially specified. In this paper, we set $ts$ to 10 as the default.
	
	Aiming at the alignment of session behaviors, we develop a Contextual LSTM (CLSTM)~\cite{ghosh2016contextual} to learn users' present consumption motivations. After the initialization, at $j$-th interaction step, the hidden state $h_{j}$ of each interaction is updated by the previous hidden state $h_{j-1}$, the current item embedding $v_{j}$, and the current behavior vector $b_{j}$ as:

	\begin{equation}
	\begin{split}
	i_{j}&=\delta(W_{vi}v_{j}+W_{hi}h_{j-1}+W_{ci}c_{j-1}+W_{bi}b_{j}+\widehat{b}_i),  \\
	f_{j}&=\delta(W_{vf}v_{j}+W_{hf}h_{j-1}+W_{cf}c_{j-1}+W_{bf}b_{j}+\widehat{b}_f),  \\
	c_{j}&=f_{j}c_{j-1}+i_{j}tanh(W_{vc}v_{j}+W_{hc}h_{j-1}+W_{bc}b_{j}+\widehat{b}_c), \\
	o_{j}&=\delta(W_{vo}v_{j}+W_{ho}h_{j-1}+W_{co}c_{j}+W_{bo}b_{j}+\widehat{b}_o),  \\
	h_{j}&=o_{j}tanh(c_{j}),
	\end{split}
	\label{CLSTM}
	\end{equation}
	
	\noindent where $i_{j}, f_{j}$ and $o_{j}$ are the input gate, forget gate and output gate at $j$-th step respectively, $v_{j}$ is the embedding item vector, $b_{j}$ is the behavior vector, $c_{j}$ is the cell memory, $\widehat{b}$ is the bias term, and $h_{j}$ is the output at $j$-th step.
	
	We essentially use the final out state $h_{t-1}$ as the representation of the present consumption motivations of user $u$, i.e., $\Psi_{SBL}=h_{t-1}$. With the above network structures, SBL can naturally model fluctuations of users' session behaviors to obtain representations of present consumption motivations.
	
	\subsubsection{Preference Behaviors Learning}
	As mentioned above, a smarter recommender system should not only consider users' present consumption motivations but also take into account the historical stable preferences. Therefore, in addition to exploiting motivations by SBL in a short term, PBL is used to learn users' stable historical preferences from the preference behaviors in a long term. Actually, only part of behaviors imply users' preferences. Thus, for determining whether a certain interaction $(v_i, b_i)\in{S_{u}}$ would be a possible element of the preference behaviors, the discrimination $D_{PBL}$ can be defined as:
	
	\begin{equation}
	D_{PBL}(v_i,b_i)=\Phi{(b_i\in{P})},
	\end{equation}
	
	\noindent where $P$ is the preference behavior set which contains the types of preference behaviors i.e., collect, cart and purchase.
	
	Different from SBL, PBL is a global representation of historical preferences with less fluctuations. That may make the architecture of SBL can not work well to obtain users' historical preferences. Inspired by Bidirectional RNN~\cite{schuster1997bidirectional}, we adapt the CLSTM to a bidirectional architecture, named Bi-CLSTM, to make full use of the contextual long-term representation in both forward and backward directions. Specifically, the cell of Bi-CLSTM is the same as Eq.~\eqref{CLSTM} and it can principally be trained with the same algorithms as a regular unidirectional CLSTM. At each time step $s$ of PBL, the forward layer with hidden state $\overrightarrow{h_{s}^{P}}$ is computed based on both the previous hidden state $\overrightarrow{h_{s-1}^{P}}$ and the current item-behavior pair $(v_{is}, b_{is})$; while the backward layer updates hidden state $\overleftarrow{h_{s}^{P}}$ with the future state $\overleftarrow{h_{s+1}^{P}}$ and the current item-behavior pair $(v_{is}, b_{is})$. Therefore, each PBL hidden representation ${h_{s}^{P}}$ can be calculated with the concatenation of the forward state and backward state, i.e., ${h_{s}^{P}}=concatenate(\overrightarrow{h_{s}^{P}}, \overleftarrow{h_{s}^{P}})$.
	
	After that, we can generate the unified representation of preference behaviors $\Psi_{PBL}$ for user $u$ through an average pooling layer:
	
	\begin{equation}
	\Psi_{PBL}=average(h_{1}^{P}, h_{2}^{P}, ..., h_{m}^{P}),
	\end{equation}
	
	Particularly, taking embedding preference interactions as inputs of above networks, PBL is able to learn and depict the profile of each user. That can help BINN to make a good understanding of users' historical stable preferences. 
	
	So far, from Discriminative Behaviors Learning (DBL), we have modeled two behavior alignments: session behaviors learning $\Psi_{SBL}$ and preference behaviors learning $\Psi_{PBL}$. Then, after an fully connected layer, we can generate the $d$-dimensions representation $\widehat{v}_{t}$ of the next item.
	
	\subsection{Model Learning and Test Stage} \label{section:ts}
	
	Taking embeddings of sequential items as inputs of networks, DBL is able to learn both users' present motivations and historical preferences by controlling recurrent states update of the two network architectures. After DBL, we can generate the prediction of next item representation $\widehat{v}_{t}$, which is a $d$-dimensions vector. In the global learning stage, we use Mean Squared Error (MSE) loss~\cite{wu2017sequential} function to learn two behavior alignments jointly from the whole set of sequential interactions $H$, which can be defined as:
	
	\vspace{-0.1in}
	\begin{equation}
	\mathcal{L}={{\frac{1}{|H|}}\sum_{S_u \in{H}}{\frac{1}{(|S_u|-ts-1)}}\sum_{t=ts+1}^{|S_u|}{\zeta(\widehat{v}_{t},{{v}_{t}})}},
	\label{loss}
	\end{equation}
	
	\noindent where $\zeta$ is MSE function, ${v}_{t}$ is the item representation that the target user $u$ is access in the next visit, $ts$ is the controlling indicator, $|S_u|$ denotes the length of the interaction sequence $S_u \in{H}$ and $|H|$ is the number of users.
	
	The Eq.~\eqref{loss} is minimized using Adagrad optimization~\cite{duchi2011adaptive}. More details of settings will be specified in experiments.
	
	So far, we have discussed the whole training stage of BINN. After obtaining the trained BINN model, in the testing stage, given an individual interaction history $S_{u}=\{(x_1, b_1), (x_2, b_2), ..., (x_{t-1}, b_{t-1})\}$, we could predict item $x_{t}$ that user $u$ is most likely to access in next visit by the following steps: (1) apply model BINN to fit user interaction process $S_{u}$ to get the user's states $\Psi_{SBL}$ and $\Psi_{PBL}$ at $t-1$ step for prediction; (2) generate the next-item embedding vector $\widehat{v}_{t}$ and compute the similarities to all the item candidates in the latent space which we have obtained in Section~\ref{section:embedding}; (3) then we can recommend the top-k potentially preferred items in the unified representation space.
	


	\section{Experiments}
	
	\begin{figure}[tbp]
		\centering
		\includegraphics[width=2.7in]{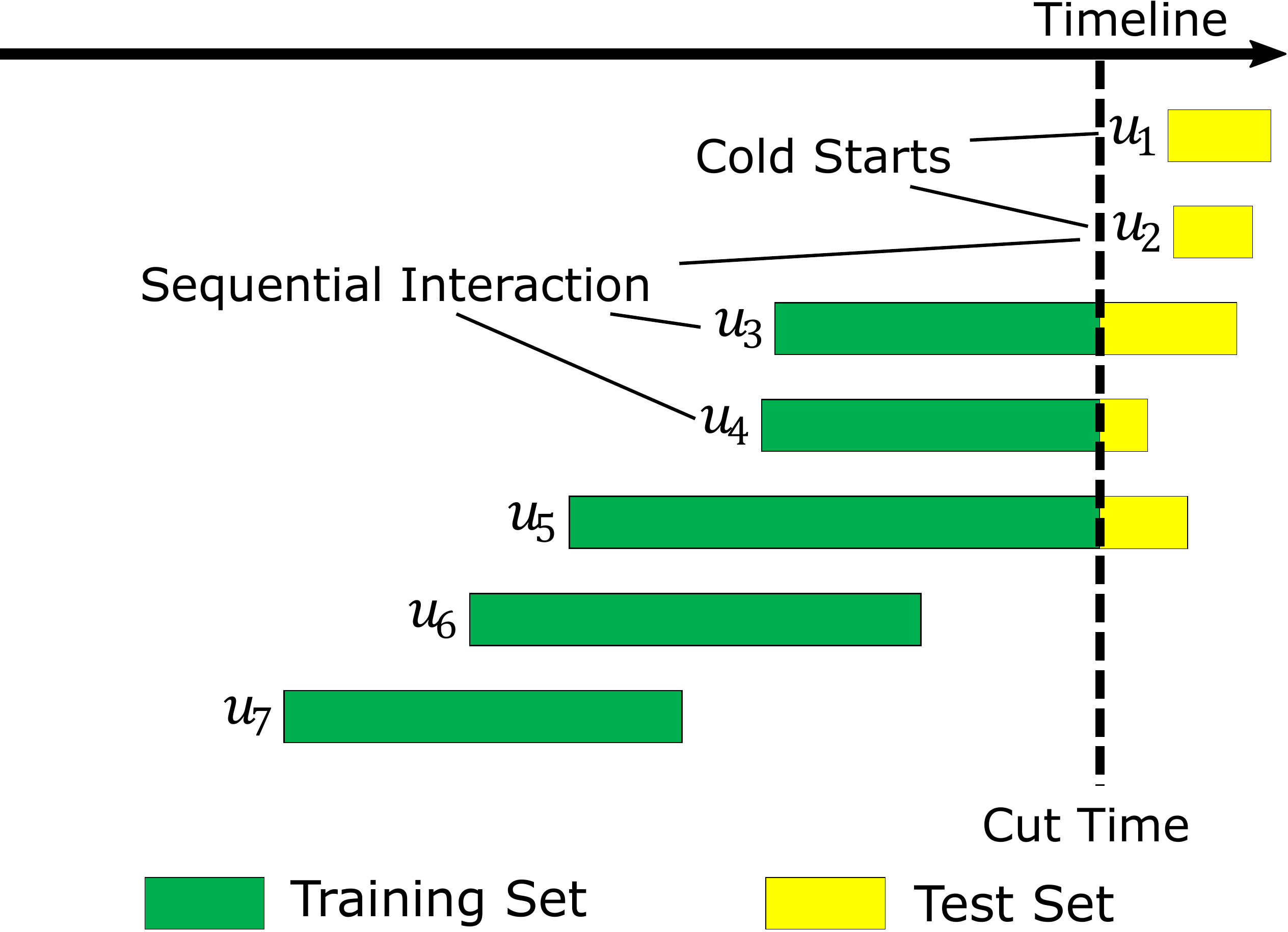}\vspace{-0.1in}
		\caption{Dataset divided with a cut time.}
		\label{dividing}\vspace{+0.1in}
	\end{figure}
	
	In this section, we first describe the experimental setups. Then, we demonstrate the effectiveness of proposed framework from the following aspects: (1) the visualization of embedding comparisons between w-item2vec in BINN and traditional item2vec, (2) the comparisons of overall recommendation performances, (3) the analysis on cold-start scenarios and (4) parameter sensitiveness of user interactions.

	\subsection{Datasets}
	Specifically, we conduct experiments on two real-world datasets, i.e., Tianchi dataset and JD dataset.
	\begin{itemize}[leftmargin=*,itemsep=2.5pt]
		\item \textbf{Tianchi}\footnote{https://tianchi.aliyun.com/getStart/information.htm?spm=5176.100067.5678.2.30a8b6d
			933N6Rr\&raceId=231522} is a public dataset by Alibaba's competition of \emph{Ali Mobile Recommendation Algorithm}, which is based on the real users-commodities behavior data on Alibaba's M-Commerce platforms. It provides 23,291,027 interactions of 20,000 customers on 4,758,484 items within a month. In this dataset, customer behaviors include click, collect, cart and purchase, the corresponding values are 1, 2, 3 and 4, respectively. 
		\item \textbf{JD} is provided by a Chinese e-commerce company Jingdong\footnote{https://www.jd.com/}, which is one of the top two largest B2C online retailers in China by transaction volume and revenue. Specially, it provides 37,087,895 interactions of 105,180 customers on 28,710 items within 75 days based on the real log data of users-commodities behaviors. In this dataset, customer behaviors include browse, cart, delete-to-cart, purchase, collect and click with the corresponding values 1, 2, 3, 4, 5 and 6, respectively. 
	\end{itemize}

	For the reliability of experimental results, we make the necessary preprocessing as follows. First, we filter the users whose interaction lengths are less than 10 and items that appear less than 5 times. Then we respectively divide two datasets into training sets and test sets according to cut time, where 90\% interactions are chosen into the training set and the remaining interactions are used for testing. Figure~\ref{dividing} illustrates strategy of dataset partitioning. In particular, for Tianchi dataset, we use 27 days data for training and the rest 3 days as test set while for JD dataset, we use 68 days data for training and the rest for test. Considering that collaborative filtering methods can not recommend an item which has not appeared before~\cite{hidasi2015session}, we filter out interactions from test set where items do not appear in the training set. In the same way, we also remove users from test set who do not appear in the training set, but we take special use of this part for analysis on the cold-start scenarios. The statistics of two datasets after preprocessing are shown in Table~\ref{dataset}.
	
	\begin{table}[tbp]
		\centering
		\caption{Statistics of datasets after preprocessing.}\vspace{-0.1in}
		\label{dataset}
		\begin{tabular}{cccc}
			\hline
			\textbf{Statistics}             & \textbf{Tianchi} & \textbf{JD}  \\ \hline
			\# of users                      &19,502                 &102,683                            \\
			\# of items                      &674,326                &24,744                        \\
			\# of behaviors                  &8,799,573              &37,061,992                          \\
			\# of behavior types             &4             		 &6                         \\
			Avg. behaviors per user          &451.30              	 &360.94                          \\
			Avg. behaviors per item          &13.05                  &1,497.82                    \\
			\# of behaviors in training set  &7,874,102              &31,811,364                            \\
			\# of behaviors in test set      &925,471                &5,250,646                         \\
			\hline
		\end{tabular}\vspace{+0.1in}
	\end{table}

	\subsection{Baseline Methods}
	We compare BINN with three traditional methods (i.e., S-POP, BPR-MF, Item-KNN) and
	two state-of-the-art RNN-based models (i.e., GRU4Rec, HRNN) each of which contains two specific implements (i.e., GRU4Rec, GRU4Rec Concat and HRNN Init, HRNN All).
	
	\begin{itemize}[leftmargin=*,itemsep=2.5pt]
		\item \textbf{S-POP} recommends the item with the largest number of interactions by the target user. This method works well in the context with high repetitiveness, and the recommendation list changes along with user interactions.
		
		\item \textbf{BPR-MF~\cite{rendle2009bpr}} is one of widely used matrix factorization methods, which optimizes a pairwise ranking objective function via stochastic gradient descent.
		
		\item \textbf{Item-KNN~\cite{linden2003amazon}} selects the items which are similar to the previously accessed items to users.
		
		\item \textbf{GRU4Rec~\cite{hidasi2015session}} uses the basic GRU with a TOP1 loss function and session-parallel minibatching.
		
		\item \textbf{GRU4Rec Concat~\cite{hidasi2015session}} is similar with GRU4Rec. Differently, we do not use the session-partition, and the users' interaction sequences are fed to the GRU4Rec independently as a whole.
		
		\item \textbf{HRNN Init~\cite{Quadrana:2017:PSR:3109859.3109896}} is a hierarchical RNN for personalized cross-session recommendations, which is based on GRU4Rec and adds an additional GRU layer to model information across the user's sessions for tracking the evolution of the user's interest. It is a state-of-the-art method in next-item recommendations.
		\item \textbf{HRNN All~\cite{Quadrana:2017:PSR:3109859.3109896}} is similar with HRNN, but the user representation generated by an additional GRU layer is used for initialization and propagated in input as each step of the next session.

	\end{itemize}
	
	For fair comparisons, we set all the hidden units in the RNN-based models as 100, their dropout probabilities and learning rate as 0.1. The embedding vector for each item is 64-dimensional in BINN model. The BINN and all the compared methods are defined and trained on a Linux server with two 2.20 GHz Intel Xeon E5-2650 v4 CPUs and four Tesla K80 GPUs.
	
	
	\subsection{Evaluation Metrics}
	As recommender systems can suggest few items each time, and the relevant items should be ranked first in the recommendation list. We therefore evaluate the personalized next-item recommendation quality with the following two evaluation metrics.
	
	\begin{figure}
		\centering
		
		\subfigure[W-item2vec.]{
			\centering
			\includegraphics[width=1.5in,height=1.45in]{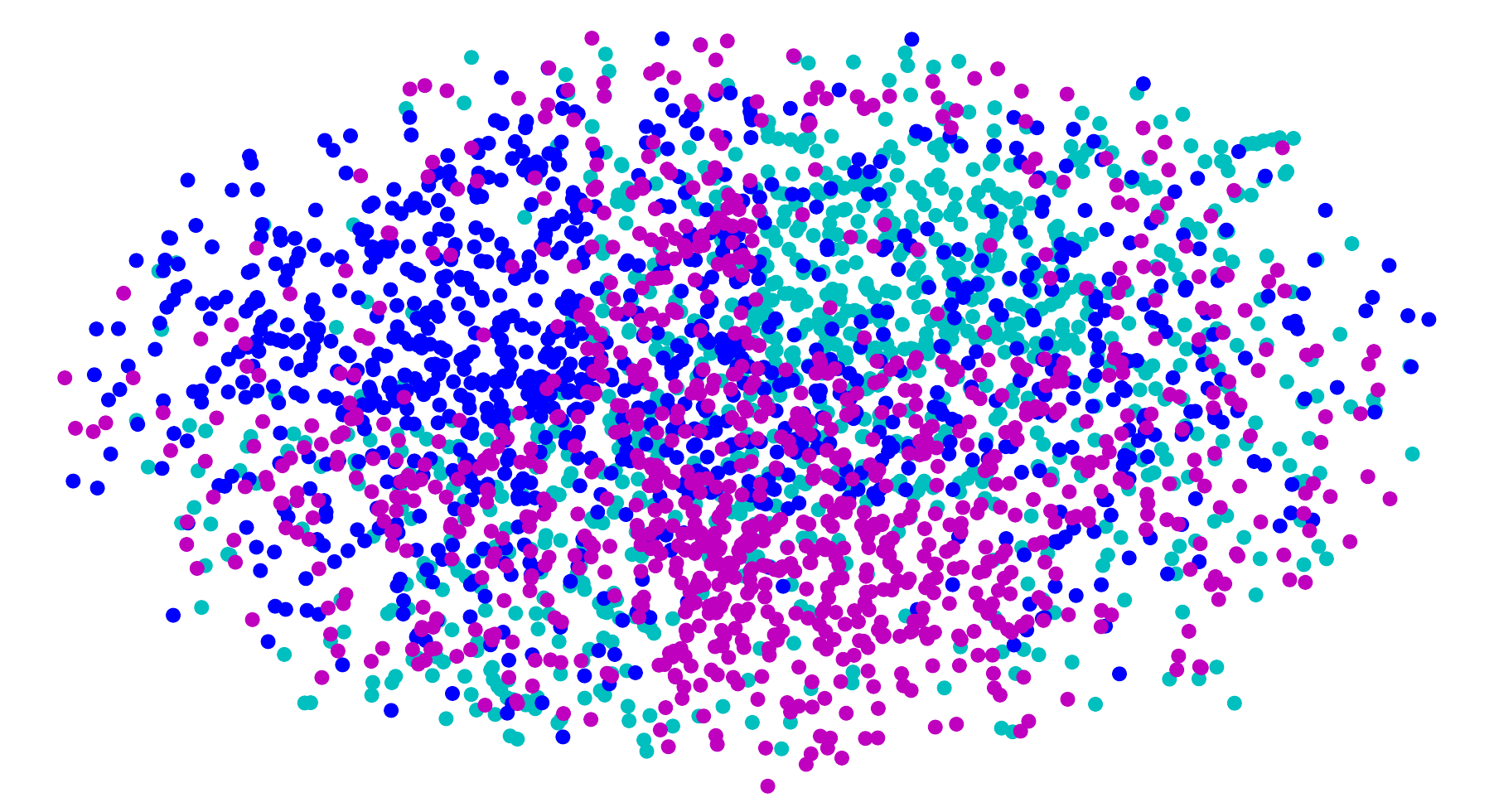}\vspace{-0.1in}
		}
		\subfigure[Item2vec.]{
			\centering
			\includegraphics[width=1.5in,height=1.45in]{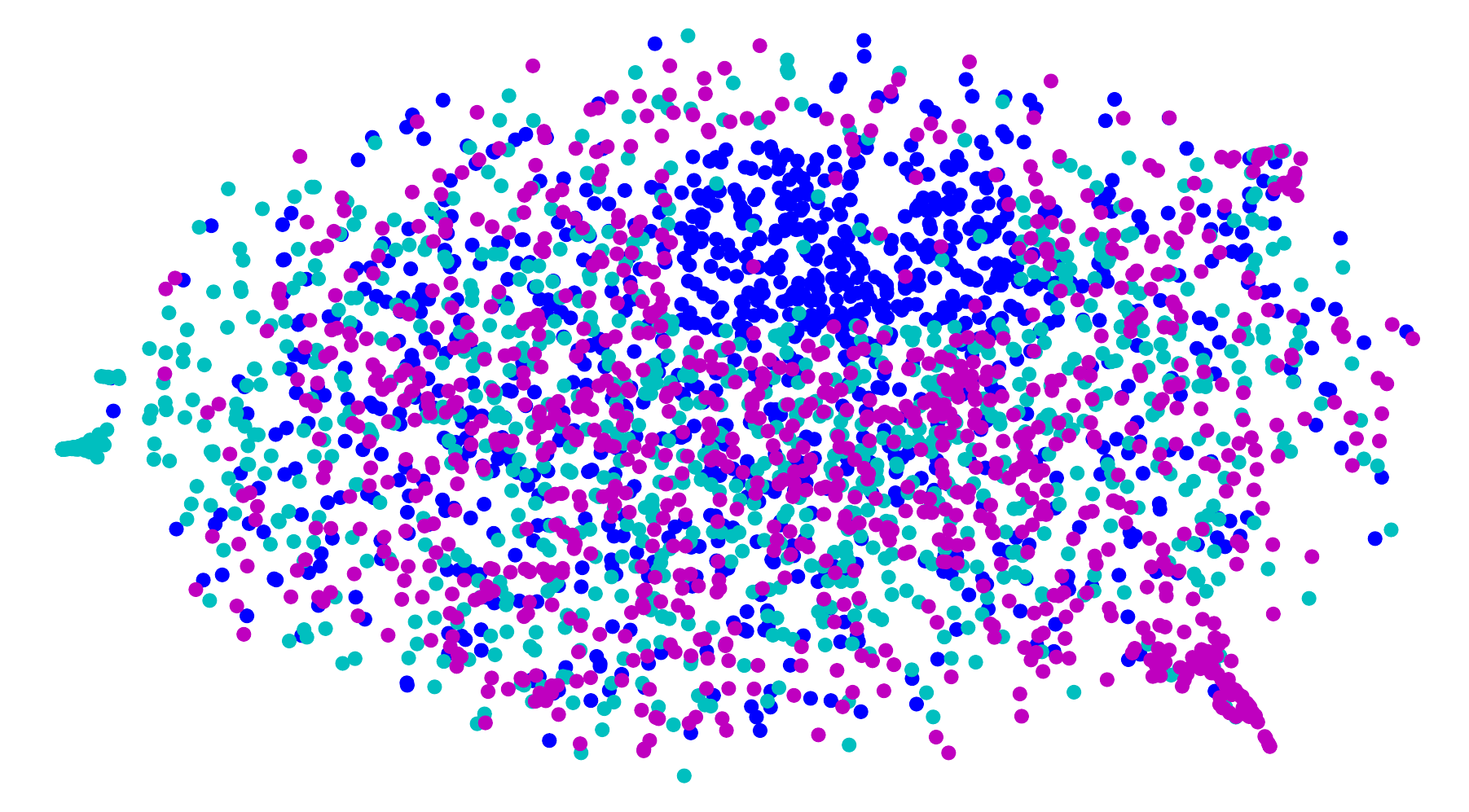}\vspace{-0.1in}
		}\vspace{-0.1in}
		\caption{T-SNE embedding for item vectors produced by w-item2vec (a), item2vec (b) on Tianchi dataset. The items are colored according to the categories.}\vspace{+0.1in}
		\label{embedd1}
	\end{figure}

	\begin{itemize}[leftmargin=*,itemsep=2.5pt]
		\item \textbf{Recall@20}. It is the primary evaluation metric that is the proportion of cases having the desired item amongst the top-20 items in all test cases~\cite{hidasi2015session, Li:2017:NAS:3132847.3132926}. Note that Recall@20 does not discriminate between items with different rankings as long as they are amongst the recommended list. In other word, the rank of items in top-20 candidate set do not make a difference. 
		
		\item \textbf{MRR@20}. Another used metric is Mean Reciprocal Rank (MRR), which is the average of reciprocal ranks of the desire items. The same with Recall metric, we set 20 as contributing value, that means the reciprocal rank is set to zero if the rank is above 20~\cite{hidasi2015session, Li:2017:NAS:3132847.3132926}. Considering the the order of recommendations matters, MRR takes the rank of each recommended item into consideration.
	\end{itemize}
	
	In summary, the higher both two evaluate metrics are, the better performances the results have.
	
	\subsection{Experimental Results}
	We first visualize the embedding of our proposed w-item2vec competing with item2vec, and then we show performances on next-item recommendation task. Finally, we discuss the cold-start problem of new users in recommendations and we analyze influences of the interaction lengths. 

	\begin{figure}
		\subfigure[W-item2vec.]{
			\centering
			\includegraphics[width=1.5in,height=1.45in]{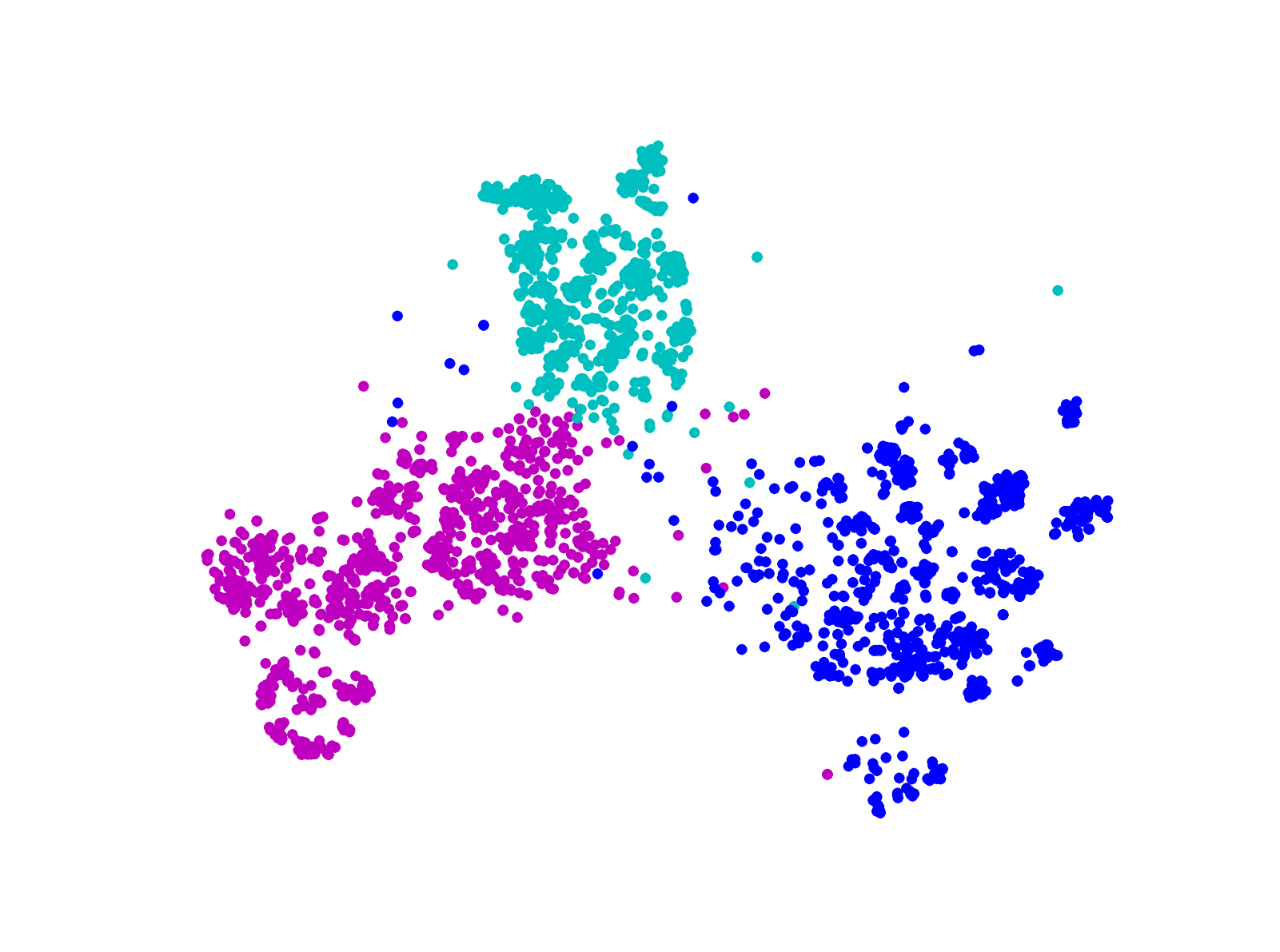}\vspace{-0.1in}
		}
		\subfigure[Item2vec.]{
			\centering
			\includegraphics[width=1.5in,height=1.45in]{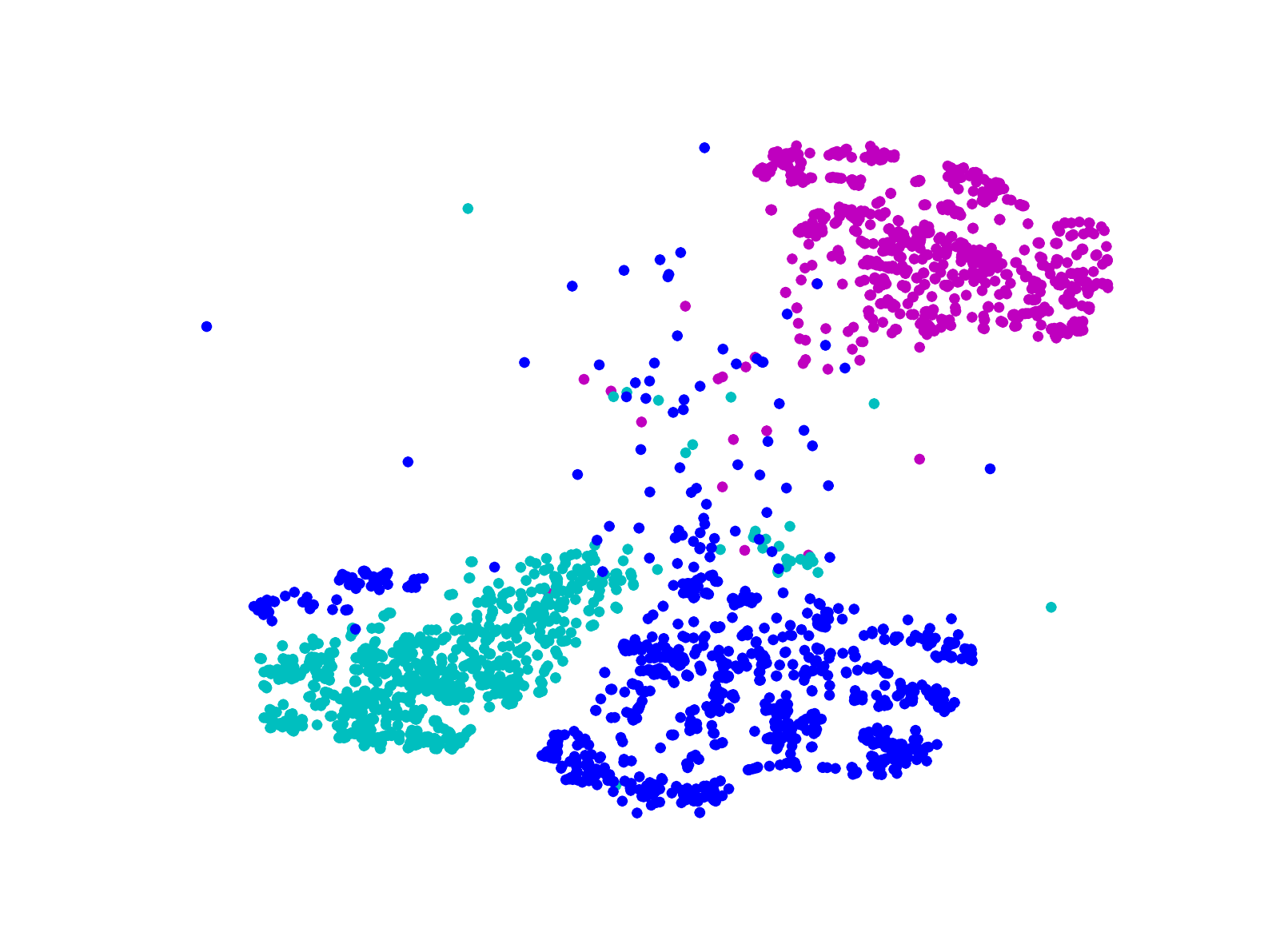}\vspace{-0.1in}
		}\vspace{-0.1in}
		\caption{T-SNE embedding for item vectors produced by w-item2vec (a), item2vec (b) on JD dataset. The items are colored according to the categories.}\vspace{+0.1in}
		\label{embedd2}
	\end{figure}

	\subsubsection{Item Embedding Visualization}
	We apply w-item2vec for generating the item embedding $v_j$ for each item $x_j$ from Eq.~\eqref{witem2vec}, in which item similarities and sequential-behavior relationships over items can be revealed simultaneously. We run the algorithm for 10 epochs and set the negative sampling value $E=10$ and compare our method with item2vec~\cite{barkan2016item2vec} on both datasets. We apply the same settings for them.
	
	Since an effective representation method can make the items implied similar attractions tend to be close to each other, we use the item categories to visualize whether the latent representations can reveal the similarities of the items. This is motivated by the assumption that a useful representation would cluster similar items in accordance with their category. To this end, we generate embeddings for 3,000 items which are randomly selected from three categories. We apply t-SNE~\cite{maaten2008visualizing} with a squared euclidean kernel to reduce the dimensionality of item embedding vectors to 2. Then we color each item point according to its category.
	
	\begin{table*}[tb]
		\centering
		\caption{Performance comparisons of BINN with baseline methods on two datasets (The improvements of RNN-based models over the best traditional method have been marked).}
		\label{Performance}
		\begin{tabular}{ccccc}
			\toprule
			\multirow{2}{*}{Methods}
			&\multicolumn{2}{c}{Tianchi}
			&\multicolumn{2}{c}{JD}\\
			\cmidrule(lr){2-3}  \cmidrule(lr){4-5}
			&Recall@20&MRR@20&Recall@20&MRR@20\\
			\midrule
			P-POP         &0.2262              &0.0824                &0.5854                 &0.2176\\
			BPR-MF        &0.0559              &0.0165                &0.1873                 &0.0664\\
			Item-KNN      &0.1964              &0.0883                &0.1246                 &0.0361\\
			
			\midrule
			GRU4Rec       &0.2025(-10.48\%)    &0.0861(-2.49\%)       &0.7034(+20.16\%)       &0.4198(+92.92\%)\\
			GRU4Rec Concat&0.2287(+1.11\%)     &0.0859(-2.72\%)       &0.7934(+35.53\%)       &0.5932(+172.61\%)\\
			HRNN Init     &0.2305(+1.9\%)      &0.0897(+1.59\%)       &0.8073(+37.91\%)       &0.6098(+180.23\%)\\
			HRNN All      &0.2167(-4.20\%)     &0.0893(+1.13\%)       &0.7762(+32.59\%)       &0.4335(+99.22\%)\\
			\textbf{BINN}&\textbf{0.2376(+5.04\%)}       &\textbf{0.0936(+6.00\%)}        &\textbf{0.8430(+44.00\%)}   &\textbf{0.7082(+225.46\%)}\\
			\bottomrule
		\end{tabular}
	\end{table*}
	
	\begin{figure*}
		\centering
		
		\subfigure[Recall@20 on Tianchi.]{
			\centering
			\includegraphics[width=1.65in,height=1.3in]{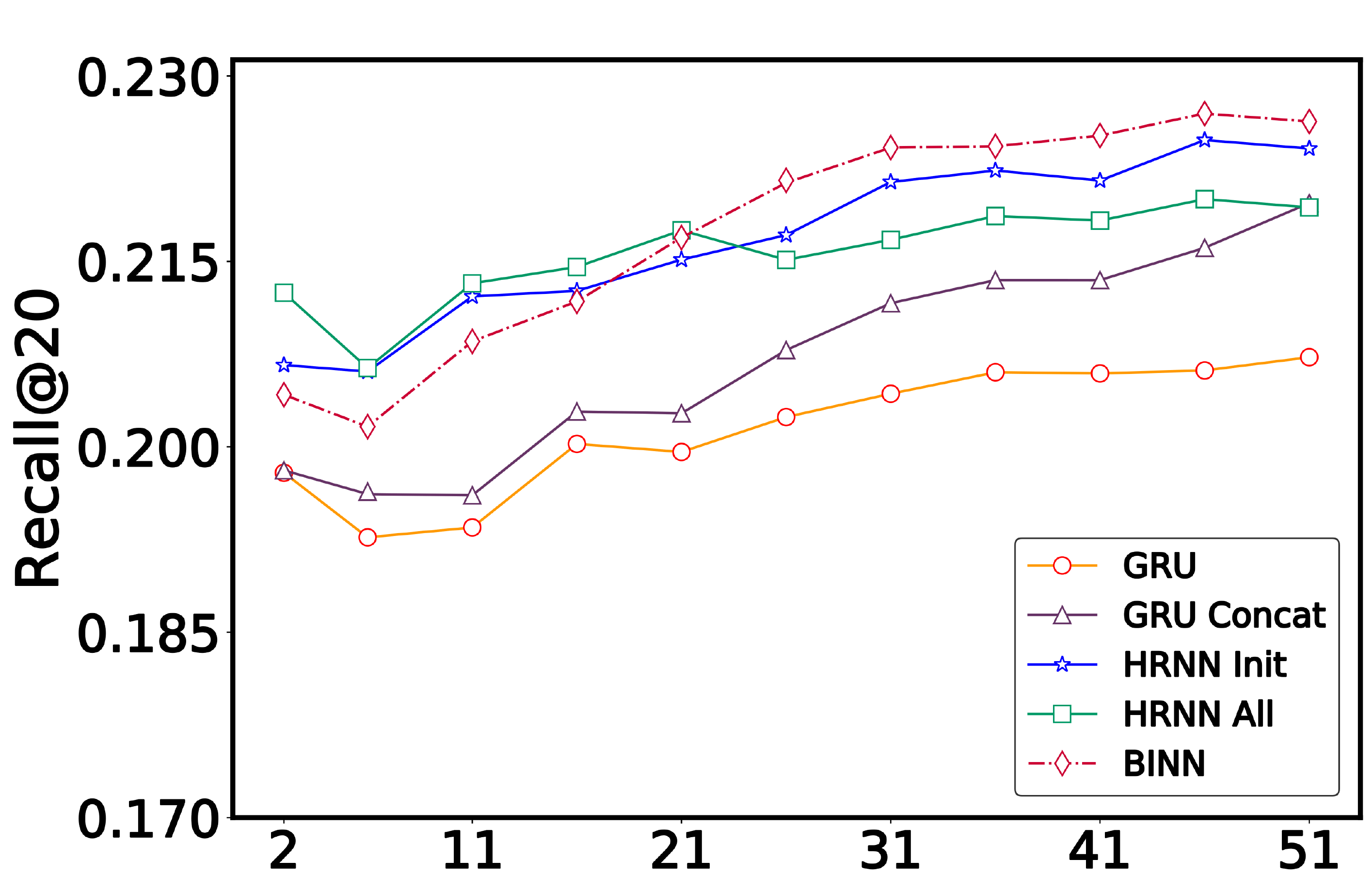}\vspace{-0.14in}
		}
		\subfigure[MRR@20 on Tianchi.]{
			\centering
			\includegraphics[width=1.65in,height=1.3in]{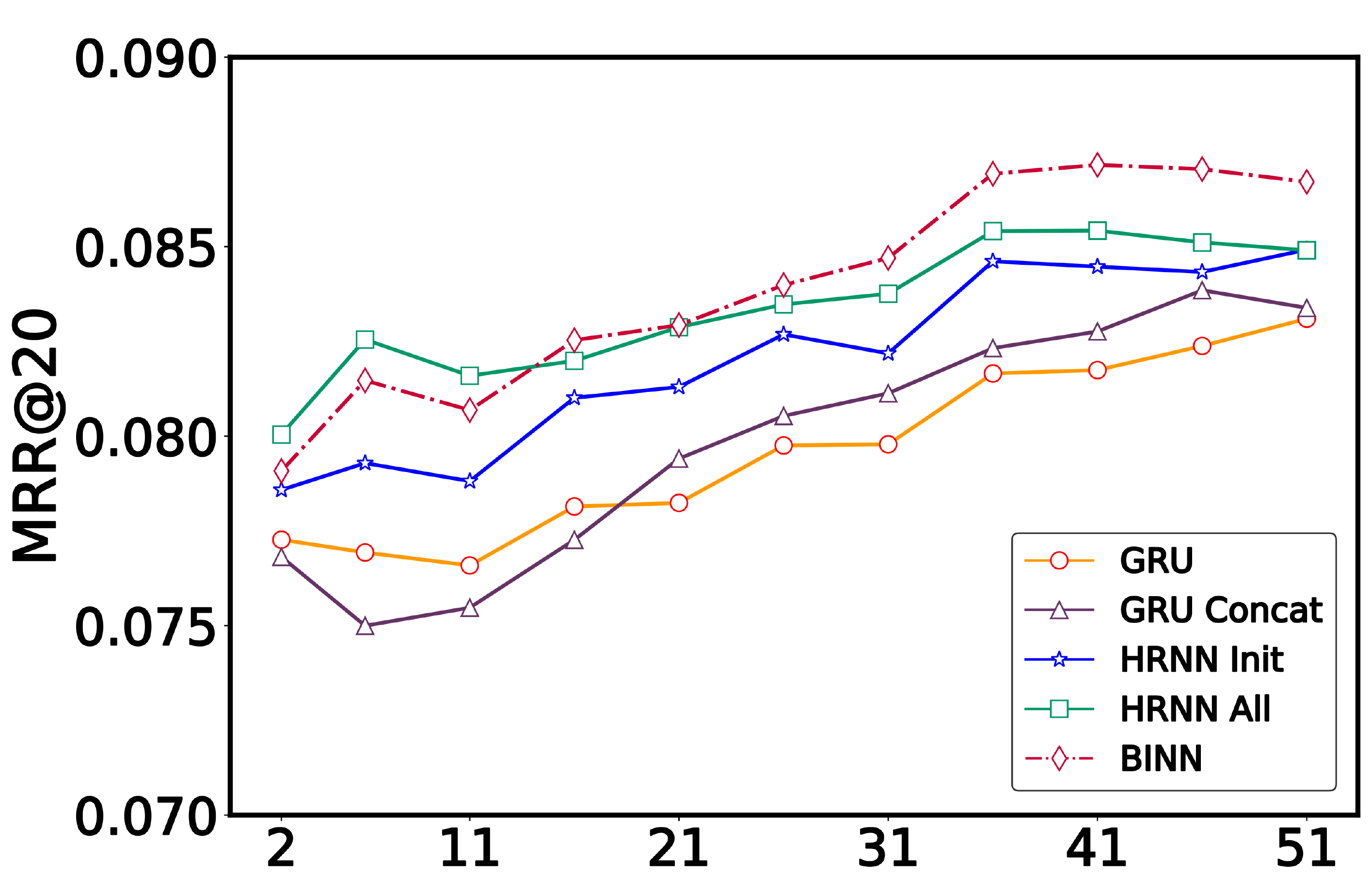}\vspace{-0.14in}
		}
		\subfigure[Recall@20 on JD.]{
			\centering
			\includegraphics[width=1.65in,height=1.3in]{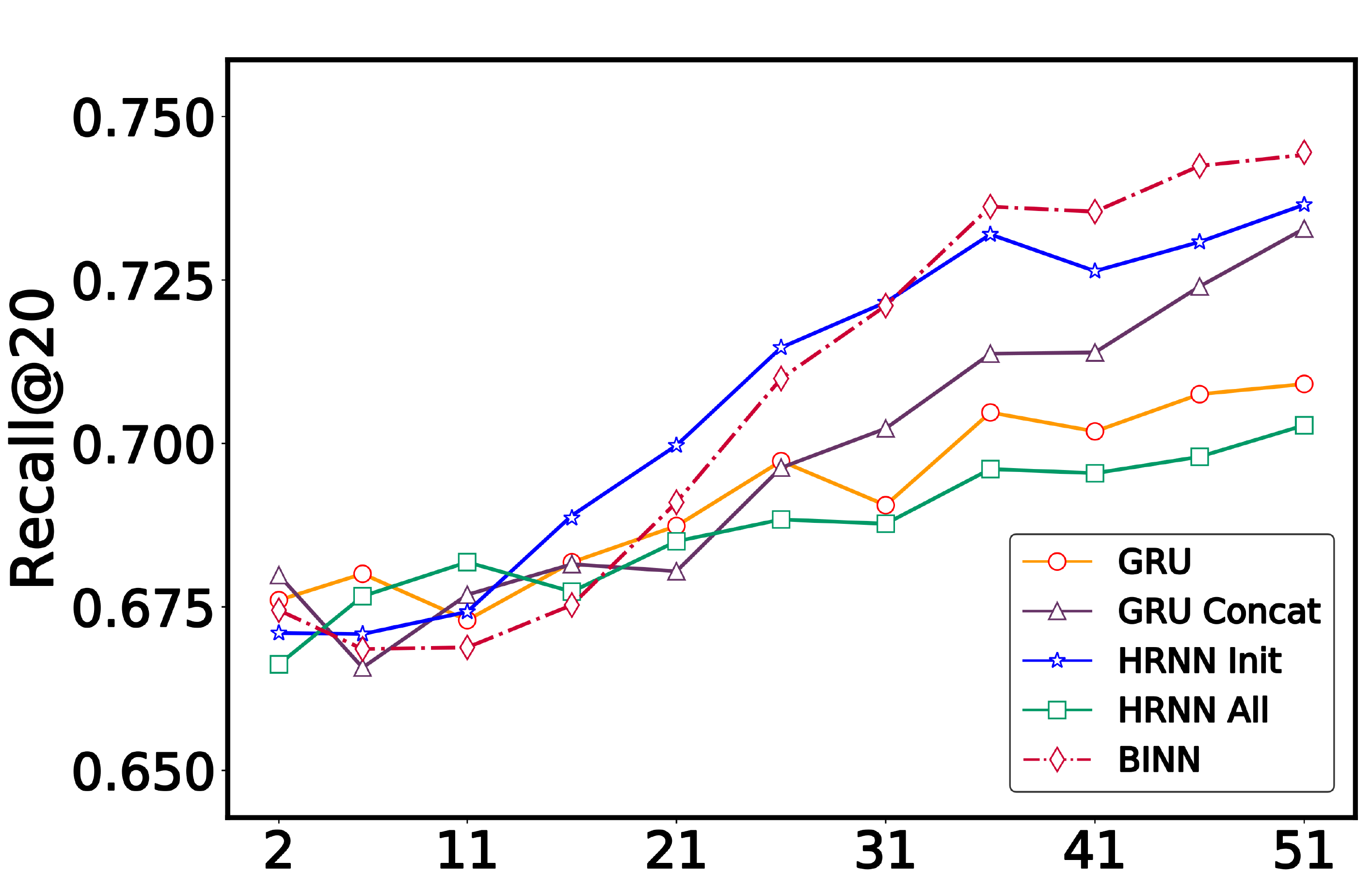}\vspace{-0.14in}
		}
		\subfigure[MRR@20 on JD.]{
			\centering
			\includegraphics[width=1.65in,height=1.3in]{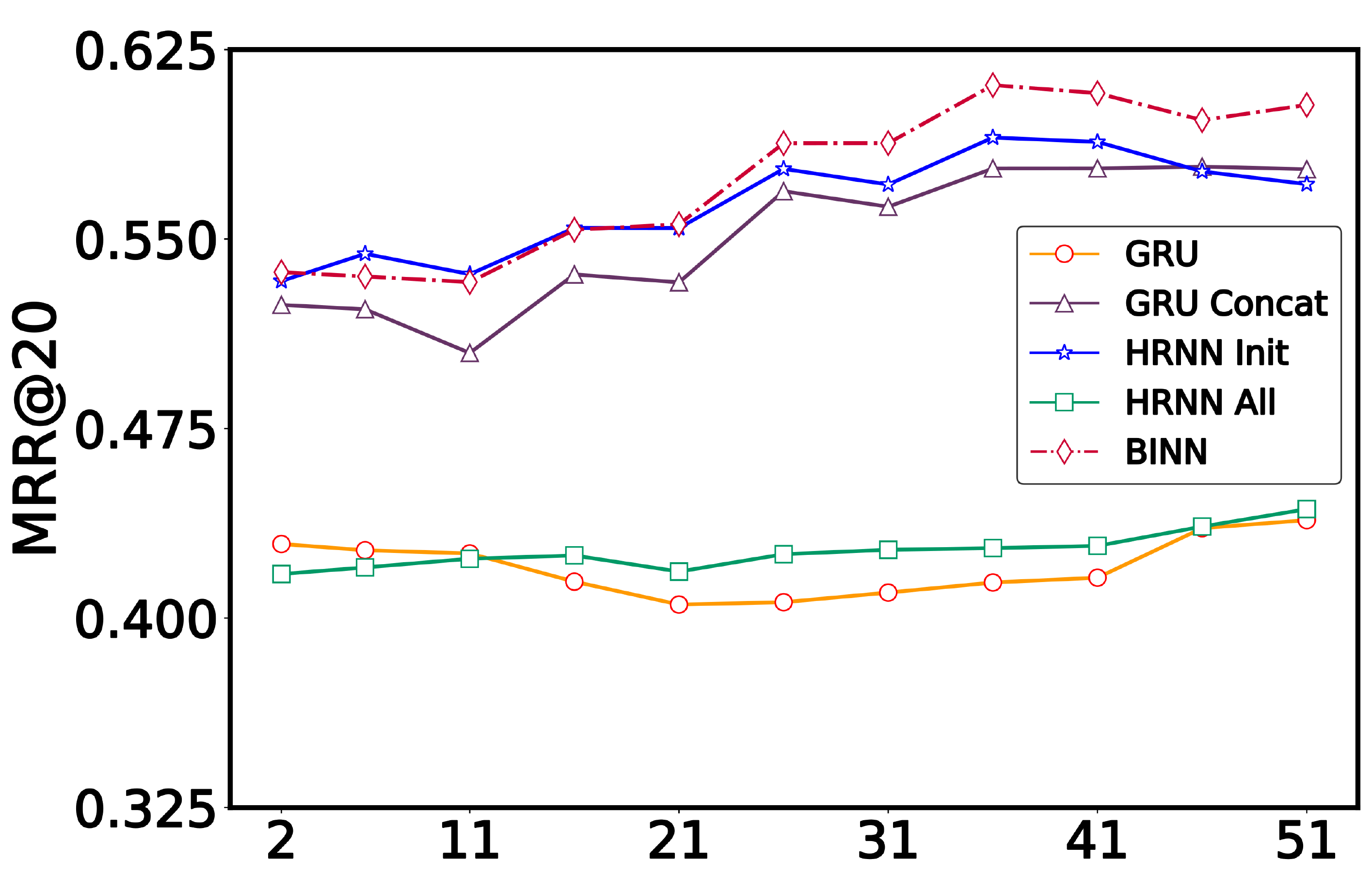}\vspace{-0.14in}
		}
		\vspace{-0.15in}
		\caption{Recommendation performances of new users cold-start on two datasets.}
		\label{coldstart}\vspace{-0.15in}
	\end{figure*}

	Figure~\ref{embedd1} and ~\ref{embedd2} present the 2D embedding that are produced by t-SNE, for w-item2vec and item2vec, respectively. As we can see, w-item2vec provides a significantly better clustering on Tianchi and also shows better performance than item2vec on JD since the clustering boundaries in Figure~\ref{embedd1}(a) and Figure~\ref{embedd2}(a) are more clear. One possible reason is that w-item2vec takes account of the item frequencies, which makes w-item2vec can generate better representation of items than item2vec. Interestingly, both two methods have shown remarkable results on JD dataset, one possible explanation could be that JD dataset is more dense. As shown in Table~\ref{dataset}, JD dataset log much more behavioral interactions on a smaller amount of items and the average behaviors per item is 1,497.82, which is much larger than Tianchi dataset 13.05 behaviors per item. That makes the model could be trained better. We further observe some outlier items that because many items on either dataset in the same category are not similar to each others.

	\vspace{-0.2in}
	\subsubsection{Recommendation Performances}
	To demonstrate the practical significance of our proposed model, we compare BINN with the other methods on the next-item recommendation task. The results of all methods on both Tianchi and JD datasets are shown in Table~\ref{Performance}. For the convenience of comparing differences between traditional and RNN-based models, we highlight the improvements of RNN-based models over the best traditional method. From the overall views, our BINN model has achieved the best performances on both two datasets.
	
	Firstly, for the results on Tianchi dataset, we have some interesting observations. BINN performs significantly better than all the other methods on both Recall@20 and MRR@20. The result indicates that BINN framework is good at dealing with personalized sequential information from the user interactions. Comparing with the RNN-based models, we can note that the traditional methods provide more competitive results. We guess a possible reason is that users' interactions on Tianchi have a high degree of repetitiveness and this dateset has a large amount of item candidates when making recommendations. That fact makes the generation of ``non-trivial'' personalized recommendations (i.e., P-POP) very challenging~\cite{Quadrana:2017:PSR:3109859.3109896}. The comparison among the five RNN-based models highlights the effectiveness to track customers' long-term preferences for next item recommendations, because models with consideration of personalized information (i.e., BINN, HRNN Init, HRNN All) outperform those methods without that (i.e., GRU4Rec, GRU4Rec Concat) on MRR@20.

	Next, we turn to the experiments on JD dataset, which exhibits some different results from those on Tianchi dataset. All the RNN-based models (i.e., GRU, HRNN and BINN) significantly outperform the traditional methods, which indicates that RNN-based models do have better abilities to model users' sequential interactions for next-item recommendations than traditional methods. In addition, the non-personalized RNN-based models GRU4Rec Concat outperforms HRNN All, which indicates that improper personalizing strategy might even make the recommendation performances worse and reveals the importance of the community trends from the short-term sequential behaviors.


	\begin{figure*}
		\centering
		\subfigure[Recall@20 on Tianchi.]{
			\centering
			\includegraphics[width=1.65in,height=1.42in]{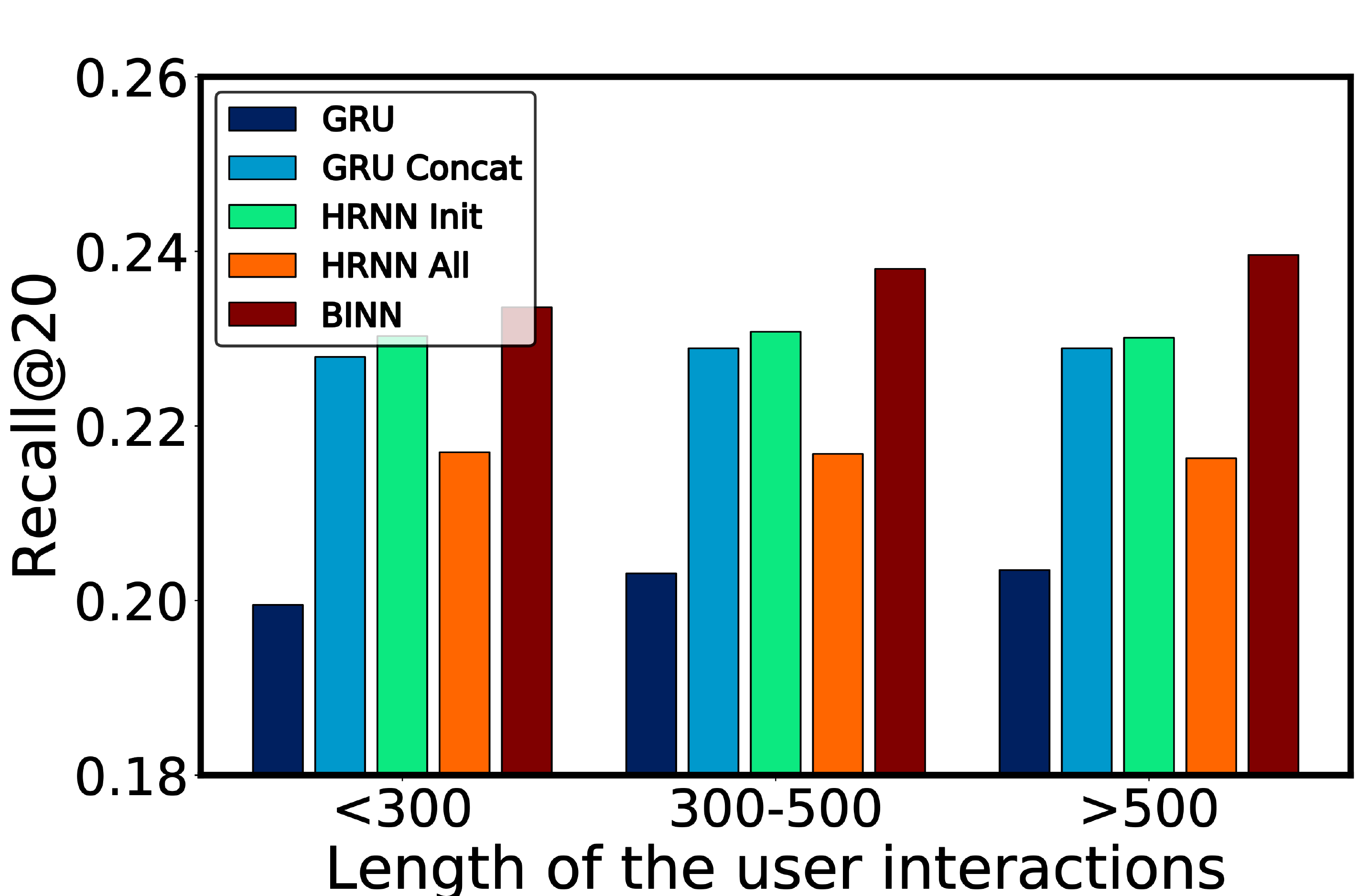}\vspace{-0.1in}
		}
		\subfigure[MRR@20 on Tianchi.]{
			\centering
			\includegraphics[width=1.65in,height=1.34in]{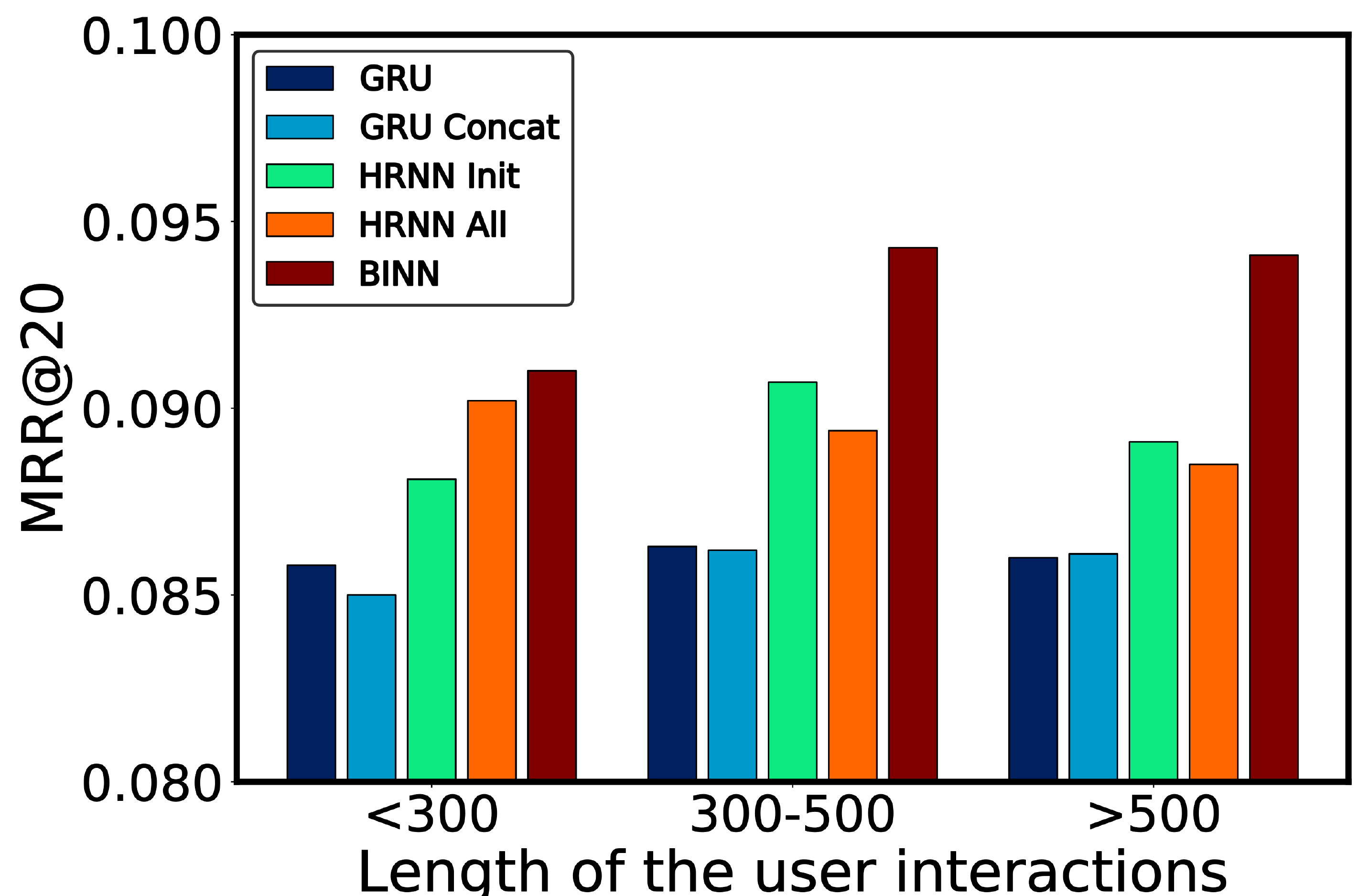}\vspace{-0.1in}
		}
		\subfigure[Recall@20 on JD.]{
			\centering
			\includegraphics[width=1.65in,height=1.34in]{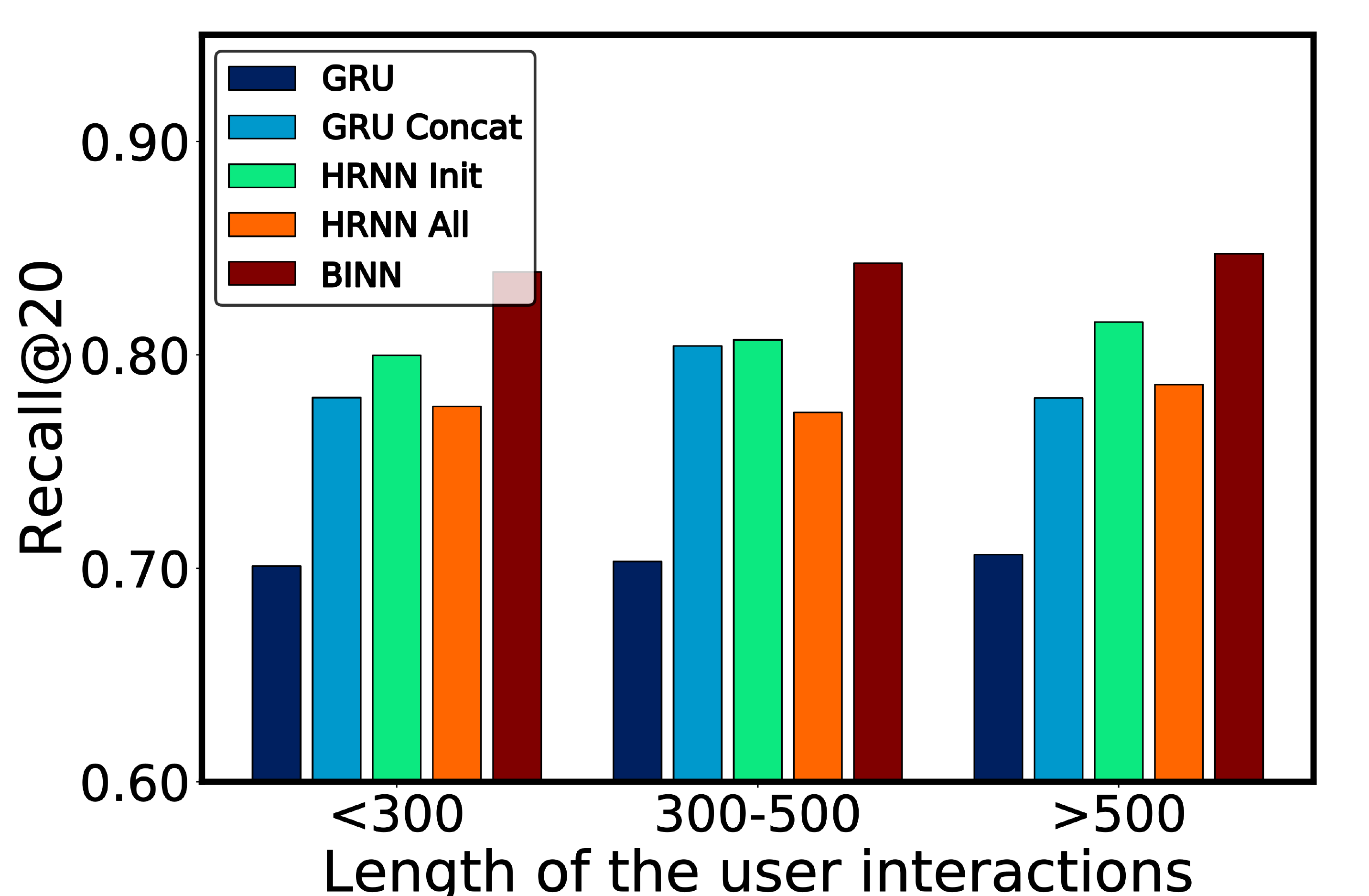}\vspace{-0.1in}
		}
		\subfigure[MRR@20 on JD.]{
			\centering
			\includegraphics[width=1.65in,height=1.34in]{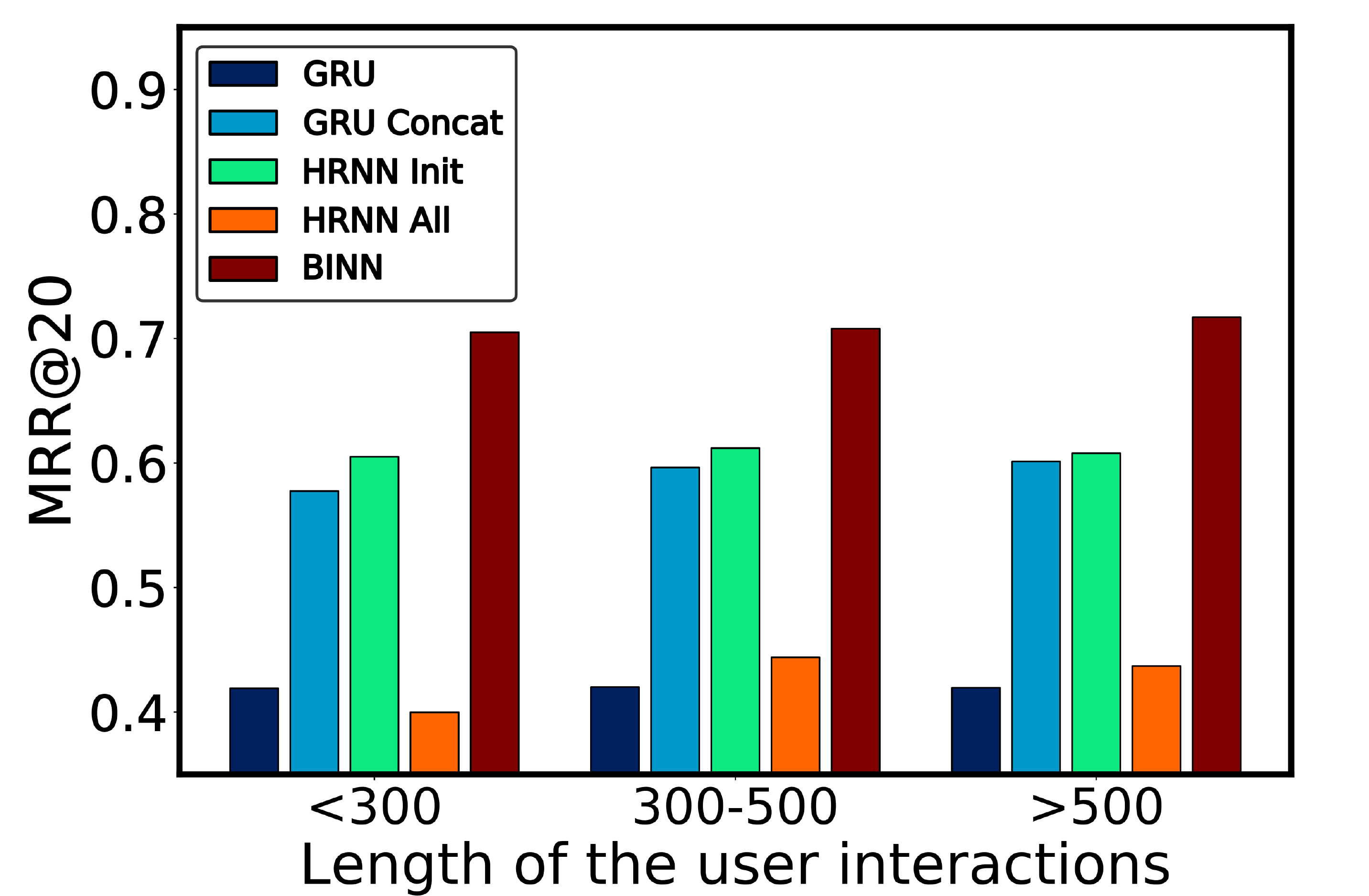}\vspace{-0.1in}
		}
		\vspace{-0.15in}
		\caption{Results of next-item recommendation over different history lengths.}
		\label{lengthChart}\vspace{-0.1in}
	\end{figure*}

	Then, we notice that RNN-based models perform much better on JD dataset than that on Tianchi dataset. One possible explanation could be that JD has more interactions of more users but less merchandises than Tianchi, and that may be a strong sequential recommendation scenario. Moreover, the number of users is much larger than the amount of items on JD dataset (i.e., the statistic of items on Tianchi dataset is 674,326, but 24,744 on JD dataset), that naturally makes the predicting candidate set smaller, and therefore, leads to more accurate result. Actually, this scenario is much common in the real world, such as online retailers and B2C platforms.
	
	In summary, BINN achieves the best performances on both two datasets, followed by HRNN Init. Both two methods take user-level representations into account. That clearly demonstrates users' interactive behaviors may follow general short-term community trends and reveal stable long-term preferences. Our BINN discriminatively models the users' historical preferences and present motivations, that leads to superior personalized recommendation quality.

	\subsubsection{Cold Start of New Users}
	Cold start is a common problem of recommender systems that new users or items have not yet gathered sufficient information to recommend or be recommended~\cite{ricci2015recommender}. As we have removed users from test set that are not in the training set on the above experiments, which is shown in Figure~\ref{dividing} shown, here we focus on these users and examine the performances of our model on cold-start problem of new users.
	
	Indeed, new users have no interactions to be pretrained and recommender systems cannot generate user profiles. That makes many user profile-based recommendation methods cannot work well, especially factorization models. However, for the RNN-based next-item recommendations, we can use a trained neural network to fit new users and predict from their second interactions one-by-one and check item rank of the next interaction. Here, we test the recommendation results on fifty items from the beginning of the second ones in the interactions of the target new users. Please note that, we do not change any training process and just select cold-start users for testing, thus all the testing do not need retraining. For better illustration, we report the results of all RNN-based models, using both metrics, respectively. 
	
	The results are shown as Figure~\ref{coldstart}. In most cases, BINN performs better than the other models on both datasets. At the beginning of the user interactions, our model BINN has deteriorated to CLSTM because of the absence of personalized preferences. Then, with the number of users' interactions growing, our model shows great improvement on recommendation performances. That can indicate the effectiveness of modeling the long-term historical preferences in BINN. What's more, all the deep learning models have shown strong capacity to face the cold-start challenges of new users. Thus, we can conclude that all the RNN-based models can work well for new users. Consequently, the results indicate the effectiveness of BINN structure.

	\subsubsection{Analysis on the User History Length}
	Here, we take further analysis on performances of our BINN model and other RNN-based models. This allows to evaluate personalized recommendation methods under different amounts of historical information and reveal the capacity of users' long-term preference representations. Since we argue the length of the user history has an impact on the recommender system performances, we divide the evaluation by the length of user interactions. Specially, we use the both datasets to make the analysis and partition users into three groups: the historical interactions less than 300, between 300 and 500, and more than 500. On account of our purpose for measuring the impact of the complex long-term preference dynamics used in BINN and other RNN-based models, we respectively record performances on these three user groups.

	Figure~\ref{lengthChart} shows the performances on both datasets. Firstly, we pay attention to improvements over the length of user behaviors grows on Tianchi dataset. We can notice that our proposed BINN has a significantly improvement on Recall@20 as the history lengths grow. Then, we turn to the results on JD dataset. For users with largest amount of interactions, our proposed BINN performs best with at least 3.92\% improvement compared to other RNN-based models. Interestingly, BINN and both two HRNN models have an improvement with the history length growing, but GRU4Rec and GRU4Rec Concat do not show continuous improvement between 300-500 and larger than 500 scales. In summary, the length of the user interactions does have an impact on the performances of recommender systems. These results clearly demonstrate the effectiveness of exploiting personalized strategies, i.e., users' historical stable preferences, to improve the recommendation performances.

	

	\section{Conclusions and Future Works}
	
	In this paper, we proposed a novel solution framework, the Behavior-Intensive Neural Network, to address the problem of personalized next-item recommendations. As a user's behaviors naturally form a interactive sequence over time, the user's historical preferences from the long-term view and present motivations from the short-term view can be dynamically revealed. Along this line, we first introduced a w-item2vec method to generate item representations by considering the sequential similarities of the superb items. Then we discriminatively exploited user behaviors and proposed two alignments of the behaviors. Specific to each alignment, we respectively developed LSTM-based neural networks to learn personal historical preferences and present consumption motivations. Finally, we conducted extensive experiments on two industrial datasets. The experimental results clearly demonstrated the effective of our proposed model in personalized next-item recommendations.
	
	In the future, we  plan to study the impact of different types of user behaviors to generate user representations and improve the next-item recommendation even further. We also plan to investigate our model in other domain, such as advertisements.

	\begin{acks}
		This research was partially supported by grants from the National~Key Research and Development Program of China (No. 2016YFB1000904), and the National Natural Science Foundation of China (No.s 61672483 and U1605251). Qi Liu gratefully acknowledges the support of the Young Elite Scientist Sponsorship Program of CAST and the Youth Innovation Promotion Association of CAS (No. 2014299).
		
	\end{acks}

	\balance
	\bibliographystyle{ACM-Reference-Format}
	\bibliography{nextitem}
	
\end{document}